\documentclass[pra,floatfix, preprint]{revtex4}
\usepackage{graphicx}

\begin{document}

\title{Spin Diffusion in Trapped Gases: Anisotropy in Dipole and Quadrupole Modes.}
\author{W. J. Mullin$^{1}$} 
\author{R. J. Ragan$^{2}$}
\affiliation {$^{1}$Physics Department, Hasbrouck Laboratory, University of
Massachusetts, Amherst, MA 01003} \affiliation{$^{2}$Physics Department,
University of Wisconsin at Lacrosse, La Crosse, WI 54601}
\date{\today }

\begin{abstract}
\noindent Recent experiments in a mixture of two hyperfine states of trapped
Bose gases show behavior analogous to a spin-1/2 system, including
transverse spin waves and other familiar Leggett-Rice-type effects. We have
derived the kinetic equations applicable to these systems, including the
spin dependence of interparticle interactions in the collision integral, and
have solved for spin-wave frequencies and longitudinal and transverse
diffusion constants in the Boltzmann limit. We find that, while the
transverse and longitudinal collision times for trapped Fermi gases are
identical, the Bose gas shows unusual diffusion anisotropy in both dipole
and quadrupole modes. Moreover, the lack of spin isotropy in the
interactions leads to the non-conservation of transverse spin, which in turn
has novel effects on the hydrodynamic modes.

\end{abstract}

\maketitle



\section{I. INTRODUCTION}

In JILA\ experiments,\cite{JILA},\cite{JILA2} a mixture of two hyperfine
states was found to segregate by species. The theoretical explanation\cite
{Levitov}-\cite{Brad} for this behavior is based on the two states playing
the role of a pseudo-spin-1/2 system, having transverse spin waves. The
theory of these new effects is based on old ideas of the transport
properties of polarized homogeneous quantum gases of real spins, such as $%
^{3}$He gas and solutions of $^{3}$He in liquid $^{4}$He,\cite{cand},\cite
{MullJeon} transcribed to the trapped gas pseudo-spin case.

Besides spin waves, the theory for homogeneous polarized fermions or bosons
led to the prediction of anisotropic spin diffusion in the degenerate state.%
\cite{cand},\cite{MullJeon},\cite{JeonGlyde} When a spin nonuniformity
is longitudinal, that is, with a variation in the magnitude of the
magnetization, the spin diffusion coefficient is $D_{\parallel }$.  On
the other hand, in a spin-echo experiment, the magnitude of the
magnetization is uniform but it varies spatially in direction.  The
corresponding diffusion coefficient, $D_{\perp \;}$ is less than
$D_{\parallel }$ when the system is polarized and degenerate.
Experimentally this feature has been seen, but was not always in
reasonable accord with theory.\cite{cand} Further, Fomin\cite {Fomin}
has suggested the effect should not exist.  However, a recent
experiment\cite{cand} has overcome several possible experimental
objections and finds good agreement with theory.  Moreover recent
papers\cite{Mineev},\cite{MR} have presented theoretical analyses that
question the validity of Fomin's approach.

Thus it seems useful to see whether a similar difference between
longitudinal and transverse diffusion in trapped gases might provide
an alternative testing ground for this question.  However, what we
show here is that the physical possibility of having differing
interaction parameters between up-up, down-down, and up-down states
(interaction anisotropy) provides a new physical basis for anisotropic
spin diffusion for bosons even in the Boltzmann limit.\cite{footnote}
For longitudinal diffusion in the Boltzmann limit only up-down
scattering contributes.  However, in the transverse case, two spins at
differing angles approach one another, and the scattering can be
analyzed as being a superposition of, say, up-up and up-down
scattering.  In the fermion s-wave case, the up-up part gives no
contribution, and, in the Boltzmann limit, the diffusion coefficients
are identical.  In that case one must go to the degenerate limit to
see the anisotropy, which then is expected to arise because the
density of scattering states differs in longitudinal and transverse
cases.\cite {MullJeon} On the other hand, for bosons, for which both
the up-up and down-down scattering rates do contribute, we find an
anisotropy even in the Boltzmann limit, but only if the various
scattering lengths differ.  We have here the interesting effect that,
although both gases obey Boltzmann statistics, there is a macroscopic
difference between fermion and boson behavior because of quantum
statistical effects in the two-particle scattering.  Moreover in the
Bose case a new element enters: the interaction now introduces a
spin-spin relaxation mechanism (a $T_{2}$ process) so that transverse
magnetization is no longer conserved.\cite{Nikuni2} For very small
relaxation times $\tau$, the hydrodynamic relaxation rate is no longer
is linear in $\tau$ ; now it diverges.  For fermions it remains well
behaved.

Below we first use the moments method to compute the spectra of the
lowest-lying longitudinal and transverse modes. However, with that method we
obtain a transverse decay rate $\gamma _{\perp }$ that diverges as $\tau $
approaches zero, in contrast to the usual diffusive behavior where $\gamma
_{\perp }\propto \tau $. In this case it is necessary to solve the \emph{%
local} hydrodynamic equations to find the correct behavior, in which the
hydrodynamic solutions are localized at the low-density regions at the edges
of the cloud where the collision time is longer. The result is a much
smaller decay rate than that obtained with the moments method.

In helium the dipole modes were detected experimentally, but in the
trapped gases it was more convenient to look at the quadrupole
modes.\cite{JILA}, \cite{JILA2} Thus we calculate both here. We have
previously reported results on the dipolar modes.\cite{MRD}

In our previous work, Ref.~\onlinecite{MullJeon}, we derived an analog of
the Landau-Silin equation for a $2\times 2$ density operator $\widehat{n}%
_{p} $ (here acting in the pseudo-spin space), with effective mean-field
single particle energy matrix $\hat{\epsilon}_{p}$. We can write the density
and single-particle energy in a Pauli representation as $\widehat{n}_{p}=%
\frac{1}{2}\left( f_{p}\hat{I}+\mathbf{m}_{p}\cdot \hat{\mathbf{\sigma }}%
\right) $ and $\widehat{\varepsilon }_{p}=\left( e_{p}\hat{I}+\mathbf{h}%
_{p}\cdot \mathbf{\hat{\sigma}}\right) $where $\mathbf{\hat{\sigma}}$ is a
Pauli matrix, $\frac{1}{2}(f_{p}\pm m_{pz})$ give the diagonal components of
the density $n_{pi}=n_{pii}$, and $\mathbf{m}_{p}$ represents the
polarization, which in equilibrium is along the axis $\mathbf{\hat{z}}$. We
find the following approximate equation for $\mathbf{m}_{p}$: 
\begin{equation}
\frac{\partial \mathbf{m}_{p}}{\partial t}-\frac{2}{\hbar }\mathbf{h\times m}%
_{p}+\sum_{i}\left[ \frac{p_{i}}{m}\frac{\partial \mathbf{m}_{p}}{\partial
r_{i}}-\frac{\partial U}{\partial r_{i}}\frac{\partial \mathbf{m}_{p}}{%
\partial p_{i}}\right] =\mathrm{Tr}\left\{ \mathbf{\hat{\sigma}}\hat{I}%
_{p}\right\} ~  \label{mmeq}
\end{equation}
with $m_{pz}(\mathbf{r})=n_{p1}-n_{p2}$ and $n_{p12}(\mathbf{r})=n_{p21}^{*}=%
\frac{1}{2}m_{p-}(\mathbf{r})=\frac{1}{2}(m_{px}-im_{py})$. The $2\times 2$
collision integral is $\hat{I}_{p}.$ The effective mean magnetic field 
is
\begin{equation}
\mathbf{h}=\frac{\hbar \Omega _{0}}{2}\mathbf{\hat{z}}+\eta \frac{t_{12}}{2}%
\mathbf{M}
\end{equation}
where 
\begin{equation}
\hbar \Omega _{0}=V_{1}-V_{2}+\left[ \left( t_{11}-t_{12}\right)
n_{1}-\left( t_{22}-t_{12}\right) n_{2}\right]\left( 1+\eta
\right). \label{EffField}
\end{equation}
 In these $\eta $ is $1$ ($-1$) for bosons
(fermions); $\mathbf{M}(\mathbf{r})=\int d\mathbf{p}/h^{3}\;\mathbf{m}_{p}(%
\mathbf{r});$ $n_{i}(\mathbf{r})=\int d\mathbf{p}/h^{3}\;n_{p}(\mathbf{r});$ 
$V_{i}$ is the external field for species $i$; $U=\frac{1}{2}(V_{1}+V_{2})$;
and $M_{z}=n_{1}-n_{2}$. The $t$'s can be evaluated in terms of the measured
scattering lengths $a_{\alpha \beta }$ by using $t_{\alpha \beta }=4\pi
\hbar a_{\alpha \beta }/m.$

The equilibrium solution in the Boltzmann limit is 
$m_{p}^{(0)}=\mathcal{M}(\beta \hbar \bar{\omega})^{3}\exp[-\beta
(p^{2}/2m+U)]$ 
where $N$ is the total number of particles, $N_{i}$ is the number of species 
$i,$ $\mathcal{M}=N_{1}-N_{2}$ is the total magnetization, and $\bar{\omega}%
\equiv (\omega _{x}\omega _{y}\omega _{z})^{1/3}.$

We have derived the collision integral for the Boltzmann case when the
various interaction paramenters differ. Our expression agrees with the same
quantity derived in Refs.~\onlinecite{Nikuni2} and \onlinecite {Brad}, and
reduces properly to previous results if all the $t$'s are taken equal.\cite
{MullJeon},\cite{LL} We find 
\begin{eqnarray}
(\sigma |\hat{I}_{p}|\sigma ^{\prime })&=&\frac{\pi}{\hbar }\int d\mathbf{p}
_{1}d\mathbf{p}_{2}d\mathbf{p}_{3}\delta (\mathbf{p}_{1}+\mathbf{p}_{2}-%
\mathbf{p}_{3}-\mathbf{p}_{4})\delta (\epsilon_{\mathbf{p}_{1}}+\epsilon _{%
\mathbf{p}_{21}}-\epsilon _{\mathbf{p}_{31}}-\epsilon _{\mathbf{p}_{4}}) 
\nonumber \\
&& \sum_{\sigma _{2}}\left\{ -t_{\sigma \sigma _{2}}^{2}\left[ (n_{\mathbf{p}%
_{1}})_{\sigma \sigma ^{\prime }}(n_{\mathbf{p}_{2}})_{\sigma _{2}\sigma
_{2}}+\eta (n_{\mathbf{p}_{2}})_{\sigma \sigma _{2}}(n_{\mathbf{p}%
_{1}})_{\sigma _{2}\sigma ^{\prime }}\right] \right.  \nonumber \\
&& -t_{\sigma ^{\prime }\sigma _{2}}^{2}\left[ (n_{\mathbf{p}_{1}})_{\sigma
\sigma ^{\prime }}(n_{\mathbf{p}_{2}})_{\sigma _{2}\sigma _{2}}+\eta (n_{%
\mathbf{p}_{1}})_{\sigma \sigma _{2}}(n_{\mathbf{p}_{2}})_{\sigma _{2}\sigma
^{\prime }}\right]  \nonumber \\
&& +\Bigl. 2t_{\sigma \sigma _{2}}t_{\sigma ^{\prime }\sigma _{2}}\left[ (n_{%
\mathbf{p}_{3}})_{\sigma \sigma ^{\prime }}(n_{\mathbf{p}_{4}})_{\sigma
_{2}\sigma _{2}}+\eta (n_{\mathbf{p}_{3}})_{\sigma \sigma _{2}}(n_{\mathbf{p}%
_{4}})_{\sigma _{2}\sigma ^{\prime }}\right] \Bigr\}  \label{collision}
\end{eqnarray}

We will linearize the kinetic equation for $\mathbf{m}_{p}$ around the
global equilibrium value $m_{p}^{(0)}\mathbf{\hat{z}}$ and use a moment
approach to compute the spin wave and diffusive damping just as done
previously.\cite{WillNik},\cite{Nikuni2} The linearized longitudinal and
transverse equations are 
\begin{equation}
\frac{\partial \delta m_{pz}}{\partial t}+\sum_{i}\left[ \frac{p_{i}}{m}%
\frac{\partial \delta m_{pz}}{\partial r_{i}}-\frac{\partial U}{\partial
r_{i}}\frac{\partial \delta m_{pz}}{\partial p_{i}}\right] =\sum_{\sigma
}\sigma (\sigma |\hat{L}_{p}|\sigma )  \label{mlong}
\end{equation}
and 
\begin{eqnarray}
&&\frac{\partial \delta m_{p+}}{\partial t}+i\frac{\eta t_{12}}{\hbar }%
\left( m_{p}^{(0)}\delta M_{+}-M_{0}\delta m_{p+}\right) -i\Omega _{0}\delta
m_{p+}  \nonumber \\
&&+\sum_{i}\left[ \frac{p_{i}}{m}\frac{\partial \delta m_{p+}}{\partial r_{i}%
}-\frac{\partial U}{\partial r_{i}}\frac{\partial \delta m_{p+}}{\partial
p_{i}}\right] =2(2|\hat{L}_{p}|1).  \label{mtrans}
\end{eqnarray}
where $\hat{L}_{p}$ is the linearized form of $\hat{I}_{p}.$

In the next section we compute results for the monopole and dipole modes.
Experiments have detected the quadrupole modes which we study in the third
section.

\section{II. Dipole Modes}

\subsection{a. Longitudinal dipole.}

\begin{figure}[h]
\centering
\includegraphics[width=5in, height=3.62in]{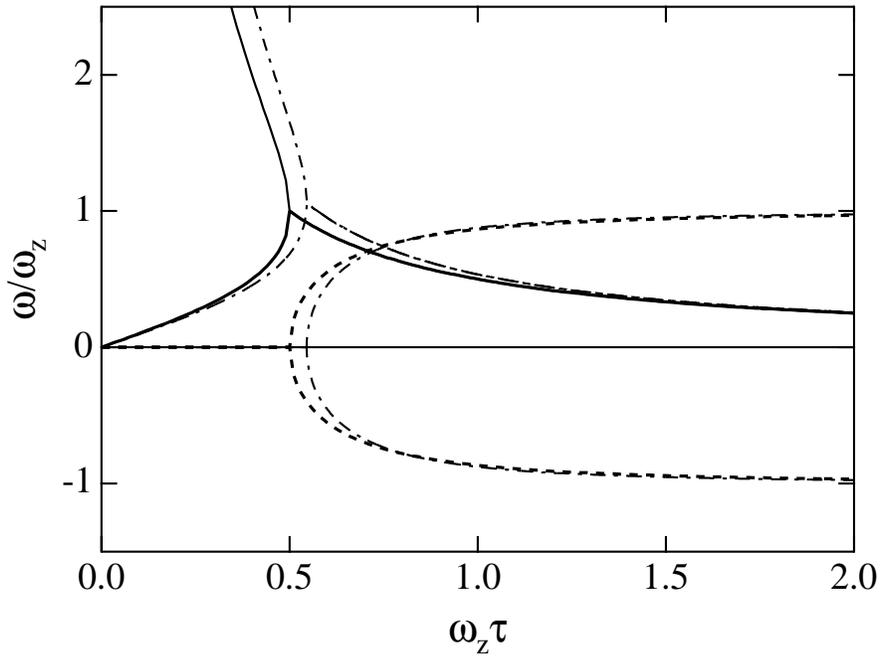}
\caption{Real (dashed) and imaginary (solid) components of longitudinal
dipole spin wave modes versus average relaxation time $\tau _{\Vert }$. Note
the linear dependence of Im$(\omega )$ for small $\tau _{\Vert }$
characteristic of diffusive behavior. The dash-dotted lines represent the
results of a numerical calculation to be discussed below.}
\label{fig:longdip}
\end{figure}

We use a variational function of the form 
\begin{equation}
\delta m_{pz}=(a_{0}+a_{1}\tilde{z}\mathbf{+}a_{2}\tilde{p}_{z}\mathbf{)}%
m_{p}^{(0)}  \label{delmform}
\end{equation}
with $\tilde{x}$ being position in units of $(\beta m)^{-1/2}\omega
_{i}^{-1} $ and $\tilde{p}_{i}$ momentum in units of $(m/\beta )^{1/2}.$ We
take the $1,$ $z,$ and $p_{z}$ moments of the kinetic equation in both the
longitudinal and transverse cases. The results for the longitudinal case, if
we assume a time dependence of $\exp (i\omega t)$ for $a_{1}$ and $a_{2},$
are 
\begin{eqnarray}
da_{0}/dt &=&0  \label{a0eq} \\
i\omega a_{1}-\omega _{z}a_{2} &=&0 \\
i\omega a_{2}+\omega _{z}a_{1} &=&-\gamma _{_{\Vert }}a_{2}  \label{LongaEqs}
\end{eqnarray}
with 
\begin{equation}
\gamma _{_{\Vert }}=\frac{4\gamma _{0}}{3}
\end{equation}
and 
\begin{equation}
\gamma _{0}=\frac{\pi \beta m^{3}\bar{\omega}^{3}t_{12}^{2}N}{h^{4}}
\end{equation}
coming from integrating the collision integral. Eq.~(\ref{a0eq}) indicates
that the monopole mode does not decay in the longitudinal case, which is
consistent with the conservation of magnetization. The effective external
field $\Omega _{0}$ does not enter into the longitudinal modes. The second
line is the magnetization equation of continuity. The relaxation rate $%
\gamma _{\Vert }$ agrees with that derived in Ref.~\onlinecite{WillNik}. To
find the modes one simply solves a quadratic equation. The dipole spectrum
is plotted in Fig.~\ref{fig:longdip} as a function of $\tau _{_{\Vert
}}\equiv 1/\gamma _{\Vert }$, the \emph{spatially averaged} collision time.
In the small $\omega _{z}\tau _{_{\Vert }}$ limit, one finds 
\begin{equation}
\omega =i\omega _{z}^{2}\tau _{_{\Vert }},
\end{equation}
which has the form of the lowest-order solution of a diffusion equation in a
harmonic potential. The other solution for small $\omega _{z}\tau _{_{\Vert
}} $ is the high frequency mode 
\begin{equation}
\omega =i\frac{1}{\tau _{_{\Vert }}}.
\end{equation}
The first of these solutions, the diffusive mode, is mainly a $z$ mode while
the second mainly a momentum dipole. For large $\omega _{z}\tau _{_{\Vert }}$
the solutions have roughly equal magnitudes of $a_{1}$ and $a_{2}$ with
frequencies 
\begin{equation}
\omega =\pm \omega _{z}+\frac{i}{\tau _{_{\Vert }}}.
\end{equation}

\subsection{b. Transverse dipole.}

We again use the form of Eq.~(\ref{delmform}). We also assume that the
effective external field $\Omega _{0}$ is quadratic in position, that is,
the part $V_{1}-V_{2}$ dominates the density dependent part in Eq. (\ref
{EffField}). We write 
\[
\Omega _{0}=\Delta _{\perp }(\tilde{x}^{2}+\tilde{y}^{2})+\Delta _{z}\tilde{z%
}^{2}. 
\]
Taking $1,$ $z,$ and $p_{z}$ moments of Eq.~(\ref{mtrans}) yields the
results 
\begin{eqnarray}
da_{0}/dt-i\omega _{0}a_{0} &=&-\gamma _{T}a_{0}  \label{DipTrans1} \\
i\left( \omega -\omega _{0}-\omega _{\Vert }\right) a_{1}-\omega _{z}a_{2}
&=&-\frac{1}{2}\gamma _{T}a_{1}  \label{cont} \\
i(\omega -\omega _{M}-\omega _{0})a_{2}+\omega _{z}a_{1} &=&-\gamma _{\perp
}a_{2}  \label{DipTrans}
\end{eqnarray}
where $\omega _{0}$ and $\omega _{\Vert }$ are parts of the effective field: 
\begin{eqnarray*}
\omega _{0} &=&2\Delta _{\perp }+\Delta _{z} \\
\omega _{_{\Vert }} &=&2\Delta _{z}
\end{eqnarray*}
\[
\gamma _{T}=\gamma _{0}(1+\eta )\sum_{\sigma }\left( \frac{t_{\sigma \sigma
}-t_{12}}{t_{12}}\right) ^{2}f_{\sigma } 
\]
with $f_{\sigma }=N_{\sigma }/N,$ and 
\begin{equation}
\gamma _{\perp }=\gamma _{_{\Vert }}\left[ \frac{7R-3S}{8t_{12}^{2}}\right]
\label{gamma2}
\end{equation}
with $R=(1+\eta )\sum_{\sigma }t_{\sigma \sigma }^{2}f_{\sigma }+(1-\eta
)t_{12}^{2}$ and 
$S=2t_{12}[(1+\eta )\sum_{\sigma }t_{\sigma \sigma }f_{\sigma }-\eta
t_{12}], $ and the mean-field frequency is 
\begin{equation}
\omega _{M}=\eta \frac{t_{12}\mathcal{M}}{\hbar }\left( \frac{\beta \hbar 
\bar{\omega}}{\sqrt{2}\lambda }\right) ^{3}
\end{equation}
where $\lambda $ is the thermal wavelength.  Since $\omega _{0}$
occurs in each equation we can remove it by including a time
dependence $e^{i\omega _{0}t}$ in each $a_{i};$ that is, we go into the
frame rotating at $\omega _{0}.$ We assume this is done and all
frequencies and will neglect this constant part.

\noindent Comments:

1) If the interactions parameters $t_{ij}$ are all equal, we have $\gamma
_{T}=0,$ $R=S=2t^{2}$ so that $\gamma _{\perp}=\gamma _{\Vert}.$ Eqs.~(\ref
{DipTrans1})-(\ref{DipTrans}) then reduce to those of Ref.~%
\onlinecite{WillNik} and the longitudinal and transverse relaxation rates
are the same, which agrees with the standard result for a homogeneous real
spin system in the Boltzmann limit.

2) For fermions, we have $\eta =-1$, so that, even if the $t$'s are \emph{not%
} equal, $\gamma _{T}=0$ and $\gamma _{\perp }=\gamma _{\Vert }.$ In
the $s-$ wave limit only up-down scattering occurs for fermions and
there can be no anisotropy in the Boltzmann limit.

3) For bosons with unequal $t$'s, the spatial averaged transverse relaxation
rate is not generally the same as the longitudinal. Moreover, we have a $%
T_{2}$-type relaxation rate for $a_{0}$ and in the equation of continuity (%
\ref{cont}). The interaction anisotropy behaves something like a
dipole-dipole interaction allowing relaxation of the transverse spin, an
effect noted previously in Ref.~\onlinecite{Nikuni2}.

If, for now, we take $\gamma _{T}=0,$ then the lowest mode in the
hydrodynamic limit $(\omega _{z}\tau _{_{\Vert }}\ll 1$) takes the form 
\begin{equation}
\omega=\omega_{\Vert}+\frac{\omega _{z}^{2}(i-\bar{\omega}_{M}\tau
_{\perp }\mathcal{)}\tau _{\bot }}{\left[ 1+(\bar{\omega}_{M}\tau _{\perp }%
\mathcal{)}^{2}\right] }.  \label{LRfreq}
\end{equation}
where $\tau _{\bot }\equiv 1/\gamma _{\perp },$ and $\bar{\omega}_{M}=\omega
_{M}-\omega _{_{\Vert }}.$ The so-called ``spin-rotation parameter'' from
Leggett-Rice systems is $\mu =\bar{\omega}_{M}\tau _{\perp }/\mathcal{M}.$ 
The form of Eq.~(\ref{LRfreq}) is the hydrodynamic frequency as modified by
spin rotation. \cite{cand},\cite{MullJeon},\cite{LL}. The first term in the
fraction is the effective diffusion frequency while the second is the
dipole-mode pseudo-spin-wave frequency. This diffusive mode is now shifted
by the effective external field, $\omega _{\Vert}$. This mode is the $%
z $ dipole mode; whereas the $p$-mode is found to be 
\begin{equation}
\omega=\omega _{M}+\frac{i}{\tau _{\bot }}
\end{equation}
which again diverges as $\tau _{\bot }$ gets small.

For large $\omega _{z}\tau _{\bot }$ we find the pair of frequencies 
\begin{equation}
\omega =\pm \omega _{z}+\frac{1}{2}\left( \omega
_{M}+\omega_{\Vert}\right) +\frac{i}{2\tau _{\bot }}.
\end{equation}

\begin{figure}[h]
\centering
\includegraphics[width=5in]{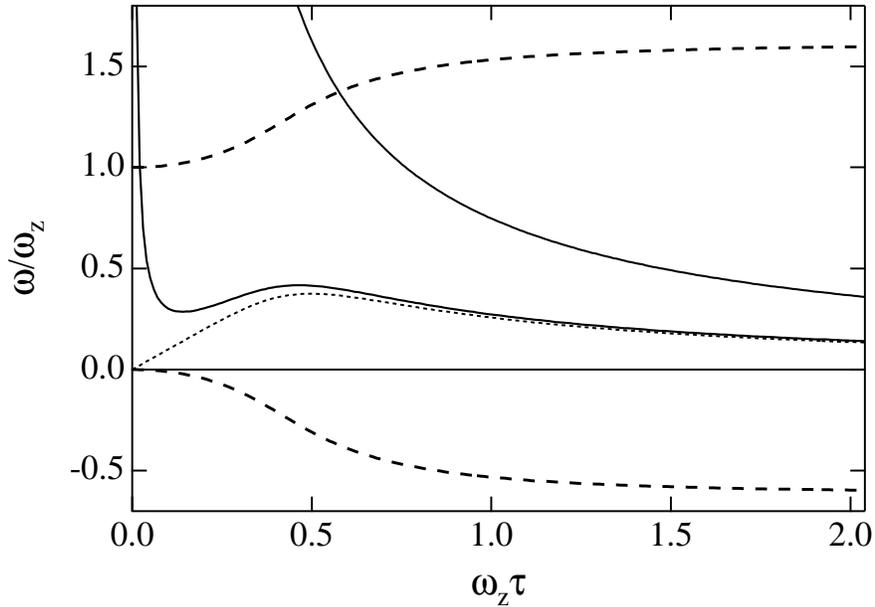}
\caption{Real (dashed) and imaginary (solid) components of transverse
dipole spin wave modes versus average relaxation time $\tau_{\perp}$
when the transverse spin is not conserved.  The dotted line shows the
lower imaginary mode when the transverse decay rate $\gamma_{T}=0$.
In this plot $\omega_{M}/\omega_{z}=1$, $\omega_{\Vert}=0$.
The linear behavior of Im$(\omega)$ for small $\tau_{\perp}$
characteristic of diffusive behavior is destroyed and replaced by a
divergence within the moments method used here.}
\label{fig:transdip}
\end{figure}

The effect of non-zero $\gamma _{T}$ is to allow a $T_{2}$ relaxation of the
transverse spins. The results are shown in Fig.~2, where we have set $1/\tau
_{\perp }=\gamma _{\Vert}$, $\gamma _{T}=0.02/\tau _{\perp }, $ and
$\bar{\omega
_{M}}=\omega _{z}.$ In the small $\tau _{\perp }$ limit, one no longer has
the hydrodynamic decay rate approaching zero, but instead it diverges at the
origin because Im$(\bar{\omega} )\approx \left( \omega _{z}^{2}\tau _{\perp
}+\gamma _{T}\right)$, and $\gamma _{T}\sim 1/\tau _{\perp }$. However,
although suitable for finite $\omega \tau$, the moments method is inadequate
in the hydrodynamic limit. It fails because the simple forms assumed for
spatial dependence cannot adjust to the spatially dependent relaxation
rates. One must solve local equations numerically for the spatial behavior.

To obtain the hydrodynamic equations we expand the momentum distribution in
terms of Hermite polynomials 
\begin{equation}
\delta m_{pz}=e^{-\beta p^{2}/2m}\sum_{k=0}c_{k}(z,t)H_{k}(p)  \label{R1}
\end{equation}
Substituting this into the kinetic equations, neglecting terms in $\omega
_{\Vert }$, integrating over the momentum, and keeping terms lowest order in 
$\tau _{\perp }$ gives, in the transverse case,

\begin{eqnarray}
\partial_t \delta M_++\partial_z J_+ =-\gamma_T(z) \delta M_+ \\
\partial_t J_++\frac{kT}{m}\partial_z \delta M_++ \omega_z^2z \delta
M_+&+&i\omega_M(z)J_+ \approx-\gamma_{\perp}(z)J_+  \label{R4b}
\end{eqnarray}
where $\delta M_+(z,t)=c_0(z,t)$ is the nonequilibrium magnetization
density, $J(z,t)=\int d\mathbf{p}/h^3 (p/m)\delta m_{pz}= c_1(z,t)$ is the
spin current, and $\gamma_{\perp}(z)=\gamma_{\perp}(0)\exp(-\beta
m\omega_z^2 z^2/2)$. Analogous equations hold in the longitudinal case. On
the RHS of Eq.~(\ref{R4b}) the $k=1$ momentum distribution has been treated
as an eigenfunction of the linearized collision integral. This is justified
by a numerical calculation of the matrix elements of the collision integral,
which gives 
\begin{eqnarray}
&&L_{\perp}[H_1(p)] = - \gamma_{\perp}(0)(1.000H_1(p) + 0.123H_3(p) 
\nonumber \\
&&-0.00094H_5(p)+ ...) \approx -\gamma_{\perp}(0)H_1(p)  \label{R3}
\end{eqnarray}

The eigenvalues of the hydrodynamic equations have been calculated
numerically for the dipole mode with boundary conditions $\delta M(0)=0$, $%
J(0)=1$, and $J(\infty)=0$, and the monopole mode with boundary conditions $%
\delta M(0)=1$, $J(0)=0$, and $J(\infty)=0$. For the longitudinal and
isotropic transverse cases this leads to only small corrections to the $\tau
\rightarrow 0$ part of the spectra obtained by the moments method as shown
in Fig.~\ref{fig:longdip}.

\begin{figure}[h]
\centerline{\includegraphics[width=5in]{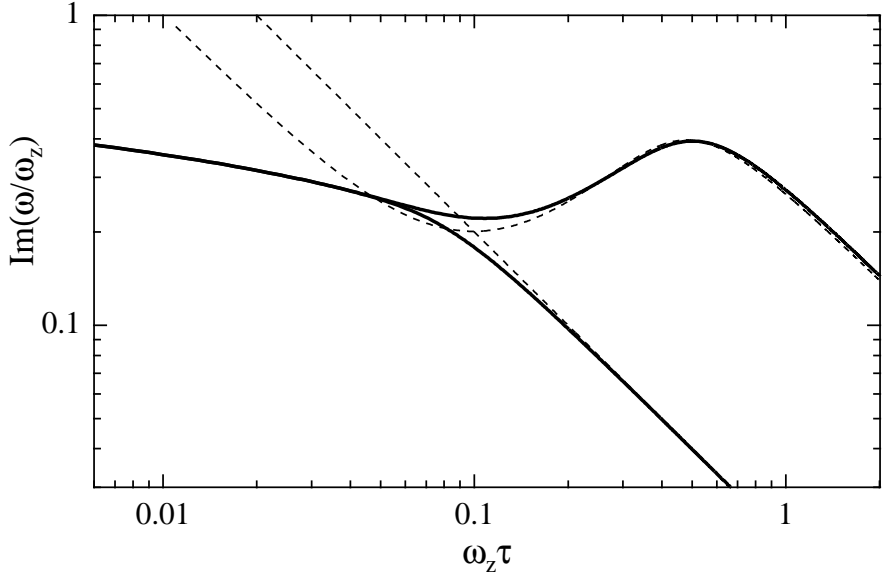}}
\caption{Imaginary part of the spin-wave spectrum vs. $\omega_z\tau_{\perp}$
for the monopole and dipole modes with $\omega_{M}= \omega_{z} $, $%
\gamma_T=0.02\gamma_{\perp}$, $\omega_{\Vert}=0$, and $\gamma_{\perp}
\approx \gamma_{\parallel}$ for both the moments method (dashed) and
hydrodynamic calculations (solid).  }
\label{fig3}
\end{figure}

\begin{figure}[h]
\centerline{\includegraphics[width=5in]{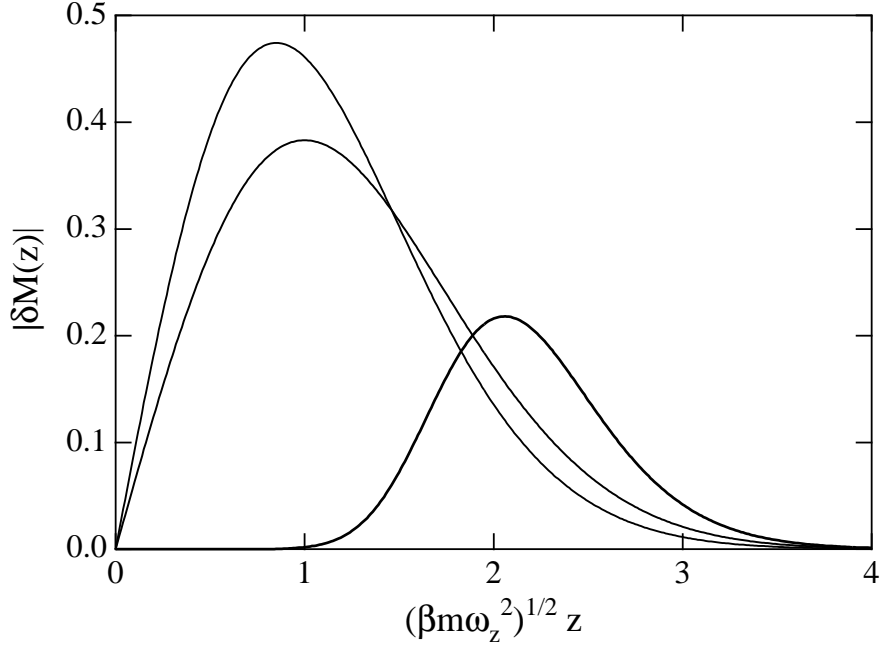}}
\caption{Profiles $|\delta M(z)|$ of the the dipole modes vs. $\protect\sqrt{%
\beta m\omega_z^2}z$, normalized with a Hermite weighting function. Tallest
peak is the hydrodynamic mode for $\gamma_T=0$ and $\tau_{\perp}=0.01$.
Middle peak is the moments method ansatz. Smallest peak is the hydrodynamic
mode for $\gamma_T=0.02\gamma_{\perp}$ and $\omega_z\tau = 0.01$. }
\label{fig4}
\end{figure}

However, for $\gamma _{T}>0$ the hydrodynamic spectrum differs
qualitatively from that of the moments method calculation.  As $\tau
_{\perp }\rightarrow 0$ the hydrodynamic dipole and monopole modes do
not decay at a rate $\sim 1/\tau _{\perp }$, but instead decay at a
slower rate.  For $\omega _{z}\tau _{\perp}\ll \alpha =
\gamma_{T}\tau_{\perp}$, the hydrodynamic equations can be Taylor
expanded about the center of an effective complex potential, giving a
decay rate
$\omega_{z}\sqrt{2\alpha\log(\sqrt{\alpha}/\omega_{z}\tau_{\perp}})$
(See Fig.~\ref{fig3}.)  At small enough $\omega _{z}\tau _{\perp }$
the $T_{2}$ decay of the magnetization at the center of the trap
causes the monopole and dipole modes to coalesce into spin-waves
localized on the lower density regions on the left and right sides of
the trap.  (See Fig.~\ref{fig4}.)

\section{III. Quadrupole Modes}

Experiments have investigated only the quadrupole modes. The increase in
theoretical complication is considerable, especially when we include unequal
scattering lengths. For the moments calculation we consider a function of
the form 
\[
\delta m_{p}=(a_{0}+a_{1}\tilde{z}^{2}\mathbf{+}a_{2}\tilde{z}\tilde{p}
_{z}+a_{3}\tilde{p}_{z}^{2}\mathbf{)}m_{p}^{(0)} 
\]
for both longitudinal and transverse cases.

\subsection{a. Longitudinal quadrupole.}

The equations we find are as follows: 
\begin{eqnarray}
a_{0} &=&-(a_{1}+a_{3}) \\
i\omega a_{1}-\omega _{z}a_{2} &=&0 \\
2\omega _{z}a_{1}+\left( i\omega +\frac{\gamma _{\Vert }}{2}\right)
a_{2}-2\omega _{z}a_{3} &=&0 \\
\omega _{z}a_{2}+(i\omega +\gamma _{3})a_{3} &=&0
\end{eqnarray}
where 
\[
\gamma _{3}=\gamma _{\Vert }\left[ 1+\frac{2}{5}(1+\eta )\sum_{\sigma }\frac{
t_{\sigma \sigma }^{2}}{t_{12}^{2}}f_{\sigma }\right] 
\]
When all the $t$'s are equal we have $\gamma _{3}\rightarrow 9\gamma _{\Vert
}/5$ for bosons and these equations reduce to those of Ref.~\cite{WillNik}.

We can again consider the solutions to the equations in limiting situations.
For very large $\omega _{z}\tau _{_{\Vert }}$ (with $\tau _{_{\Vert
}}=1/\gamma _{\Vert })$ we find the the three solutions 
\begin{equation}
\omega \cong \left\{ 
\begin{array}{l}
\pm 2\omega _{z}+\frac{i}{4}(\gamma _{3}+\gamma _{\Vert }) \\ 
i\frac{\gamma _{3}}{2}
\end{array}
\right.
\end{equation}

\begin{figure}[h]
\centering
\includegraphics[width=5in, height=5.55in]{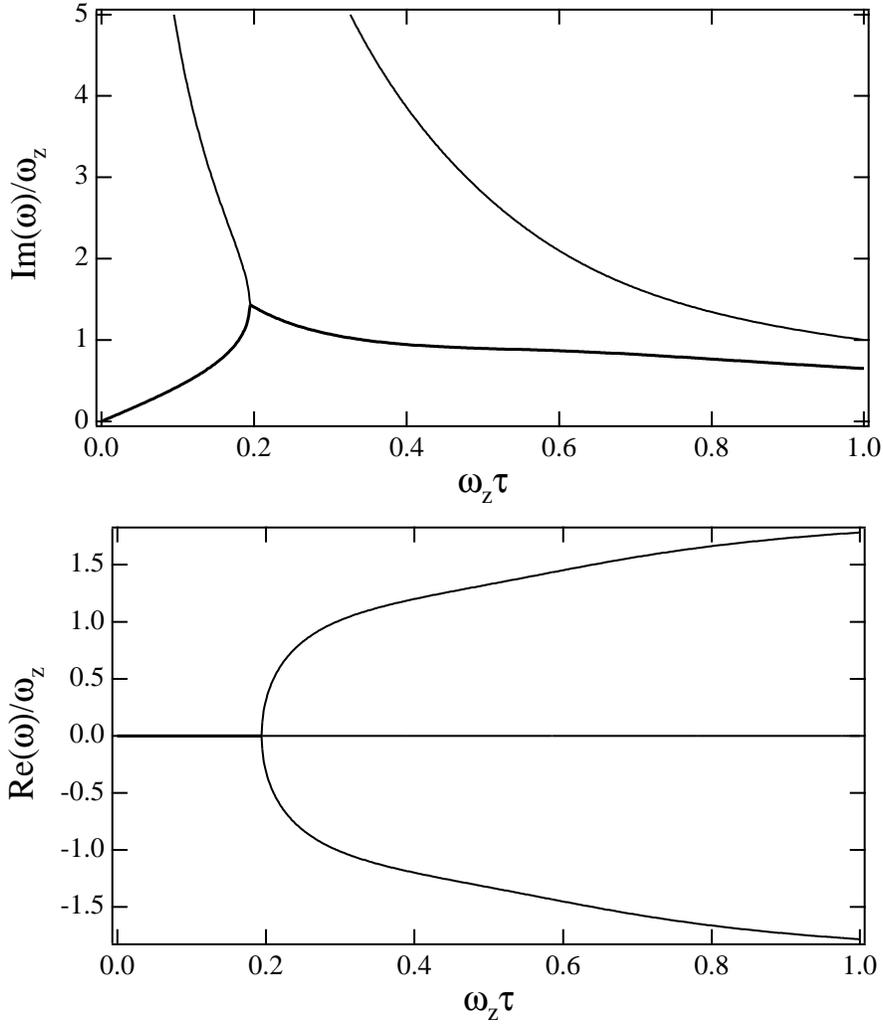}
\caption{Imaginary(top) and real (bottom) components of longitudinal
quadrupole spin wave modes versus average relaxation time $\tau _{\Vert }$. 
The diffusive mode is shown with darker lines.}
\label{fig:longquad}
\end{figure}

In the small $\omega _{z}\tau _{_{\Vert }}$ limit we again find a diffusive
mode (proportional to a second order Hermite function in $z^{2}):$%
\[
\omega =4i\omega _{z}^{2}\tau _{_{\Vert }},
\]
as well as two higher frequency decaying modes (one in $zp_{z}$ and the
other the Hermite in $p^{2}):$%
\begin{equation}
\omega \cong \left\{ 
\begin{array}{l}
\frac{i}{2\tau _{_{\Vert }}} \\ 
i\gamma _{3}
\end{array}
.\right. 
\end{equation}
Fig.~\ref{fig:longquad} gives shows the modes for all $\omega _{z}\tau _{_{\Vert }}.$

\subsection{b. Transverse quadrupole}

Taking moments of the quadrupole magnetic equation Eq.~(\ref{mtrans}) and
again going into the frame rotating at $\omega _{0}$ gives the following
equations 
\begin{eqnarray}
i\omega \left( a_{0}+a_{1}+a_{3}\right) -i\omega _{\Vert }a_{1}+\gamma
_{T}\left( a_{0}+\frac{1}{2}a_{1}+\frac{7}{6}a_{3}\right) &=&0 \\
i\omega \left( a_{0}+3a_{1}+a_{3}\right) -i\omega _{\Vert }\left(
a_{0}+6a_{2}\right) -2\omega _{z}a_{2}+\frac{\gamma _{T}}{2}\left( a_{0}+%
\frac{3}{2}a_{1}+\frac{7}{6}a_{2}\right) &=&0 \\
i\left( \omega -\frac{\omega _{M}}{2}\right) a_{2}-i\omega _{\Vert
}a_{2}+\omega _{z}(a_{1}-a_{3})+\frac{\zeta _{2}}{2}a_{2} &=&0 \\
i\omega \left( a_{0}+a_{1}+3a_{3}\right) -i\omega _{\Vert }a_{1}+2\omega
_{z}a_{2}-2i\omega _{M}a_{3}+\frac{7}{6}\gamma _{T}\left( a_{0}+\frac{1}{2}%
a_{1}\right) +\zeta _{3}a_{3} &=&0
\end{eqnarray}
where 
\begin{eqnarray}
\zeta _{2} &=&\frac{4}{3}\gamma _{0}\left\{ 1+\frac{(1+\eta )}{8}%
\sum_{\sigma }f_{\sigma }\left[ \left( \frac{t_{\sigma \sigma }-t_{12}}{%
t_{12}}\right) ^{2}+6t_{\sigma \sigma }\frac{t_{\sigma \sigma }-t_{12}}{%
t_{12}^{2}}\right] \right\} \\
\zeta _{3} &=&\gamma _{0}\left[ \frac{8}{3}+(1+\eta )\frac{77}{60}+\frac{%
(1+\eta )}{t_{12}^{2}}\sum_{\sigma }f_{\sigma }\left( \frac{79}{20}t_{\sigma
\sigma }^{2}-\frac{25}{6}t_{\sigma \sigma }t_{12}\right) \right]
\end{eqnarray}
In the case of equal values of all the $t$'s we have $\zeta
_{2}\rightarrow \gamma _{\Vert }$ and $\zeta _{3}\rightarrow 18\gamma
_{\Vert }/5$ for bosons.  As in the dipole case the first equation
represents a nonconserved monopole mode.  If we take $\omega _{\Vert
}$ and $\gamma _{T}$ to zero, this equation represents the zero
frequency monopole mode.  We see then that a non-zero $\omega _{\Vert
}$ couples the monopole and quadrupole modes, because it arises from a
quadratic external field. We will now consider solutions in various special
cases.

\emph{Zero values of }$\gamma _{T},$ $\omega _{M},$\emph{\ and
}$\omega _{\Vert }$: When all three of these parameters are set to
zero, we get modes that look just exactly like the longitudinal case
of Fig.~\ref{fig:longquad}.  The diffusive mode's imaginary part goes
to zero with $\omega _{z}\tau _{2}$ as expected; here $\tau
_{2}=1/\zeta _{2}$.  Two of the modes have degenerate relaxation times.

\begin{figure}[h]
\centering
\includegraphics[width=5in, height=5.44in]{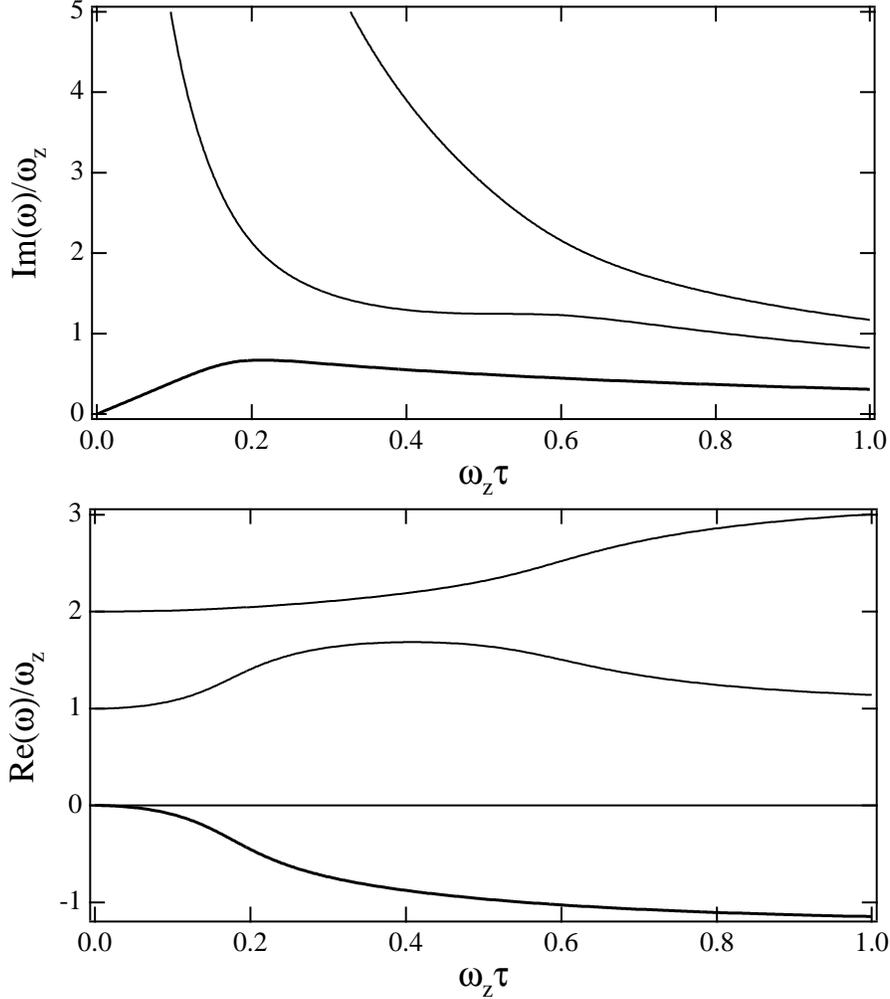}
\caption{Imaginary(top) and real (bottom) components of longitudinal
quadrupole spin wave modes versus average relaxation time $\tau _{2}$
with $\omega_{M}=2\omega_{z}$, $\gamma_{T}=0$, $\omega_{\Vert}=0$.  The
degeneracy in some imaginary parts is lifted, and the diffusive mode
(dark lines) now has a real part.}
\label{fig:MuQuad}
\end{figure}

\emph{Non-zero }$\omega _{M};$ \emph{zero }$\gamma _{T}$\emph{\ and
}$\omega _{\Vert }$: With non-zero mean-field value $\omega _{M}$ the
degeneracy is now split as shown in Fig.~\ref{fig:MuQuad}.  For small
$\omega _{z}\tau _{2},$ the real parts of two of the modes have finite
values, $\omega _{M}$ and $\omega _{M}/2,$, but the imaginary parts of
those modes still diverge so that they are very short-lived.  The
diffusive mode, whose imaginary and real parts both go to zero as
$\tau_{2}\rightarrow 0$, is given, to order ($\omega _{z}\tau _{2})^{2}$, by
\begin{equation}
\omega \cong 4\omega _{z}^{2}\tau _{2}(i-\omega _{M}\tau _{2}\mathcal{)}
\end{equation}
This behavior is of the same form as the transverse dipole mode, Eq.(\ref
{LRfreq}) with the factor of 4 occurring for this higher order mode. The
relaxation time $\tau _{2}$ reduces to $\tau _{\Vert }$ only when all the $t$%
's are equal or for fermions. So even ignoring the effect of $\gamma _{T}$
we would have boson diffusive anisotropy since $\tau _{2}$ is not the same
as $\tau _{\Vert }.$ The monopole mode decay rate vanishes in this case.

\begin{figure}[h]
\centering
\includegraphics[width=5in, height=2.86in]{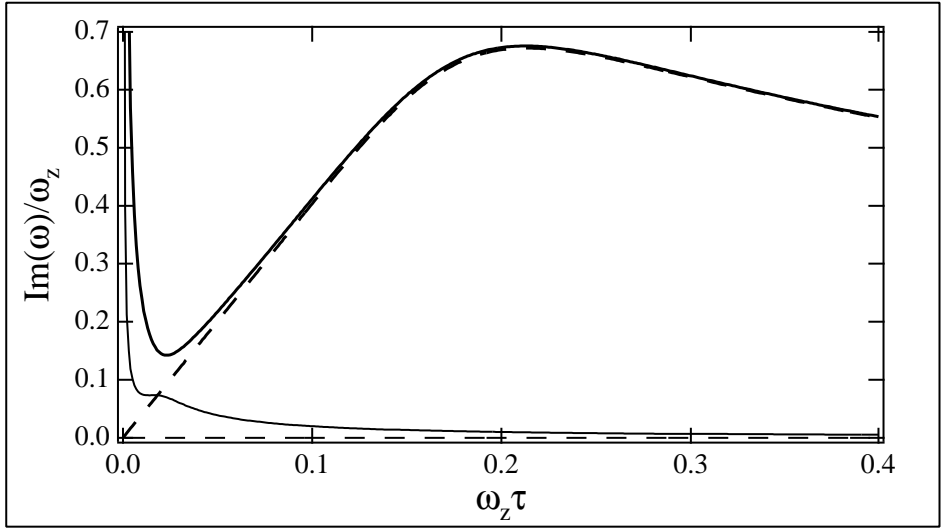}
\caption{Imaginary components of transverse monopole and quadrupole
boson spin wave modes versus average relaxation time $\tau_{2}$ with
$\gamma_{T}=0.002/\tau_{2}$ and $\omega_{\Vert}=0$.  The solid lines are the modes with
$\gamma_{T}$ non-zero and the dashed lines show the previous results
with $\gamma_{T}$ zero.  Now these modes diverge at small $\tau$.}
\label{fig:AlphaQuad}
\end{figure}

\emph{Non-zero }$\gamma _{T};$\emph{\ and zero }$\omega _{\Vert }$: As
in the dipolar case the existence of the non-conservation terms means
a different behavior for small $\tau _{2}$ as we now see in
Fig.~\ref{fig:AlphaQuad}.  From the moments calculation we find again
that the imaginary parts of \emph{all}
modes, including both monopole and diffusive, now diverge proportional to $%
\gamma _{T}$ $\sim 1/\tau _{2}.$ The figure shows just the imaginary
parts of the monopole and diffusive, indicating that the transverse
magnetization is no longer conserved for bosons.  However, as in the
dipole case, a more accurate analysis shows that the small $\tau $
behavior shows a divergence as $\sqrt{\log{(1/\omega_{z}\tau_{2})}}$.
Fermions on the other hand behave as usual, because their scattering
can depend only on $t_{12}$ and not on the difference in $t$ values.

\begin{figure}[h]
\centering
\includegraphics[width=5in, height=5.45in]{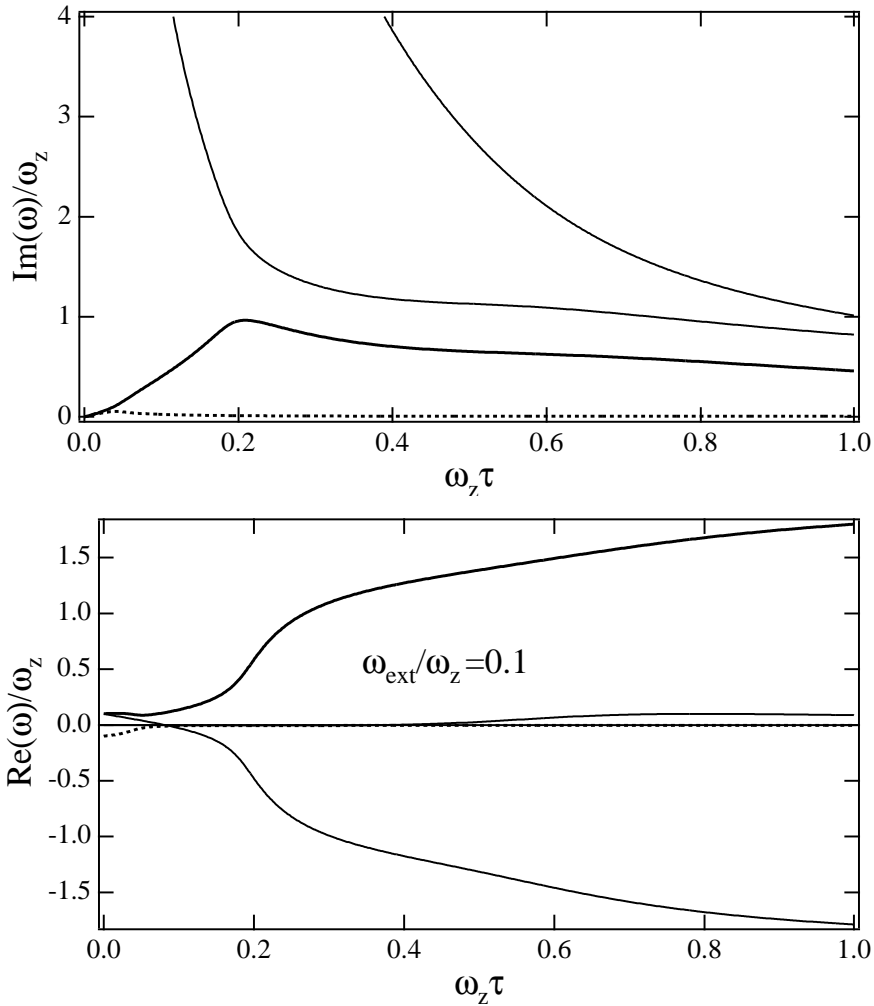}
\caption{Imaginary (top) and real (bottom) components of the
transverse monopole and quadrupole boson spin wave modes versus
average relaxation time $\tau_{2}$ with
$\omega_{\parallel}=0.1\omega_{z}$.  The darker lines are the
diffusive modes and the dotted line is the monopole mode; these are now
mixed at small $\tau_{2}$.  The very small $\tau_{2}$ behavior of the
diffusive relaxation frequency and all the real modes is altered by
the presence of this coupling with the monopole mode. See the next figure.}
\label{fig:wAll}
\end{figure}

\begin{figure}[h]
\centering
\includegraphics[width=5in, height=2.85in]{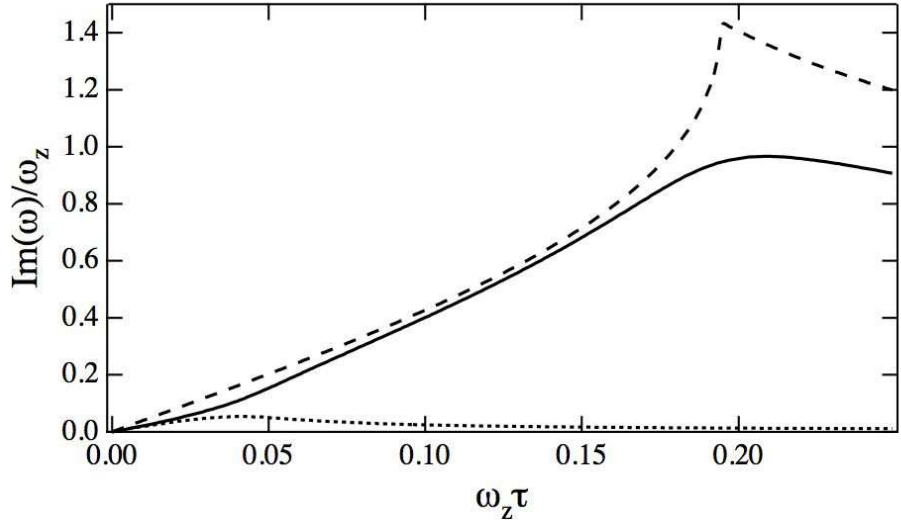}
\caption{Imaginary components of the transverse monopole (dotted
line)and diffusive quadrupole (solid line) boson spin wave modes versus average
relaxation time $\tau _{2}$ with
$\omega_{\parallel}=0.1\omega_{z}$.  The dashed lines shows
the $\omega_{\parallel}=0$ case for comparison.}
\label{fig:wDetail}
\end{figure}

\emph{Non-zero }$\omega _{\Vert };$\emph{\ zero }$\omega _{M}$ \emph{and }$%
\gamma _{T}$\emph{\ }: The quantity $\omega _{\Vert }$ arises from the
existence of an external quadratic potential difference between the
two species.  Because it is quadratic, it causes a coupling between
the monopole and quadrupole modes.  In Fig.~\ref{fig:wAll} we show the
transverse quadrupole modes with a small value of $\omega _{\Vert
}>0$.  We see that it does cause splitting between the previously
degenerate
relaxation times. Further, the real frequencies that vanished for small $%
\omega _{z}\tau _{2}$ now are now finite and split with non-trivial
structure. A particularly interesting feature for quite small $\omega
_{z}\tau _{2}$ is the near coalescing of the imaginary monopole and
diffusive modes due to the quadratic coupling. The upper mixed mode mode,
instead of going to zero as 4$\omega _{z}\tau _{2},$ as it did when
diffusive now has half that slope as demonstrated in
Fig.~\ref{fig:wDetail} .

\section{IV. Discussion}

We have shown that rather unusual differences between Bose and Fermi gases
in the Boltzmann limit can arise in spin-diffusion experiments, when the
scattering lengths for different spin configurations are not the same, and
when the effective field for the transverse modes does not vanish. For
fermions there is no distinction between longitudinal and transverse modes.
In the fermion case to see such an anisotropy one would have to go to the
degenerate limit where phase space differences would lead to a difference
between longitudinal and transverse modes.\cite{MullJeon} However, for
bosons, longitudinal and transverse modes show striking differences when the
scattering lengths are not the same. Indeed in the transverse case the
magnetization is no longer conserved as it is in the longitudinal case;
instead of a linear $\omega _{z}\tau $ dependence of the spin diffusive mode
one finds a divergence.

In experiments on Rb, the interaction anisotropy is very small. To test the
novel effects predicted here one might use Na,\cite{sodium} which has a
difference in interaction paraments; Numerically we estimate that for $^{23}$%
Na $\gamma _{\perp }$ can differ from $\gamma _{\Vert }$ by as much as 14\%
with $\gamma _{T}/\gamma _{\perp }\approx 0.04$. Interaction differences
might also be induced by using Feshbach resonance methods.

We thank Dr.\ Jean-No\"{e}l Fuchs and Prof. David Hall for useful
discussions.


\begin{thebibliography}{99}
\bibitem{footnote}  We consider temperatures such that the distribution
function is Boltzmann, but the pair interactions are quantum mechanical with
s-wave scattering predominating. This can occur if the interparticle spacing
is much greater than the de Broglie wavelength which, in turn, is much
greater than the scattering length.
\end{thebibliography}
\end{document}